\documentclass[12pt,a4paper]{article}
\usepackage{graphicx}
\usepackage{amssymb,amsfonts,amsmath}
\usepackage{cite}

\usepackage{color}

\newcommand{\green}[1]{}

\title{Evolutionary dynamics in finite populations with zealots}
\author{Yohei Nakajima${}^{1}$ and Naoki Masuda${}^{1,*}$
\\
\ \\
\ \\
${}^{1}$ 
Department of Mathematical Informatics,\\
The University of Tokyo,\\
7-3-1 Hongo, Bunkyo, Tokyo 113-8656, Japan
\ \\
${}^*$ Corresponding author: masuda@mist.i.u-tokyo.ac.jp}
\date{\today}

\begin{document}

\setlength{\baselineskip}{0.77cm}
\maketitle

\newcommand{\EQ}{Eq.~}
\newcommand{\EQS}{Eqs.~}
\newcommand{\FIG}{Fig.~}
\newcommand{\FIGS}{Figs.~}

% \begin{frontmatter}

%\author[utokyo]{Yohei Nakajima
%}
%\author[utokyo]{Naoki Masuda\corref{cor1}
%}
%\address[utokyo]{Department of Mathematical Informatics,
%The University of Tokyo,
%7-3-1 Hongo, Bunkyo, Tokyo 113-8656, Japan}
%\cortext[cor1]{masuda@mist.i.u-tokyo.ac.jp}

%\end{frontmatter}

%\newpage

\section*{Abstract}

We investigate evolutionary dynamics of two-strategy matrix games with zealots in finite populations. Zealots are assumed to take either strategy regardless of the fitness. When the strategy selected by the zealots is the same, the fixation of the strategy selected by the zealots is a trivial outcome. We study fixation time in this scenario. We show that the fixation time is divided into three main regimes, in one of which the fixation time is short, and in the other two the fixation time is exponentially long in terms of the population size. Different from the case without zealots, there is a threshold selection intensity below which the fixation is fast for an arbitrary payoff matrix. We illustrate our results with examples of various social dilemma games.

\newpage

\section{Introduction}\label{sec:introduction}

A standard assumption underlying evolutionary game dynamics, regardless of whether a player is social agent or gene, is that players tend to imitate successful others. In actual social evolutionary dynamics, however, there may be zealous players that stick to one option according to their idiosyncratic preferences regardless of the payoff that they or their peers earn. Collective social dynamics in the presence of zealots started to be examined for non-game situations such as the voter model representing competition between two equally strong opinions (i.e., neutral invasions) \cite{Mobilia2003PhysRevLett,GalamJacobs2007PhysicaA,Mobilia2007JStatMech,XieSreenivasan2011PRE,SinghSreenivasan2012PRE}.
Zealots seem to be also relevant in evolutionary game dynamics. For example,
voluntary immunization behavior of individuals
when epidemic spreading possibly occurs in a population
can be examined by a public-goods dilemma game
\cite{Fu2011ProcRSocB}. In this situation, some individuals
may behave as zealot such that they
try to immunize themselves regardless of the cost of immunization
\cite{LiuWuZhang2012PhysRevE}.

In our previous work, we examined evolutionary dynamics of 
the prisoner's dilemma and snowdrift games in infinite populations with zealots
\cite{Masuda2012SciRep}. Specifically, we assumed zealous cooperators and asked the degree to which the zealous cooperators facilitate
cooperation in the entire population.
We showed that cooperation prevails
if the temptation of unilateral defection is weak
or the selection strength is weak. For the prisoner's dilemma, we analytically obtained the condition of cooperation.

In the present paper, we conduct a finite population analysis of evolutionary dynamics of a general two-person game with zealots. Evolutionary games in finite populations have been recognized as a powerful analytical tool for understanding properties of evolutionary games such as conditions of cooperation in social dilemma games. In addition, the outcome for finite populations is often different from that for infinite populations
\cite{Nowak2004Nature,Taylor2004BullMathBiol,Nowak2006book}. We take advantage of this method to understand evolutionary dynamics of games with zealots for general matrix games.

It should be noted that the fixation probability, i.e., the probability that a given strategy eventually dominates the population as a result of stochastic evolutionary dynamics, is a primary quantity to be pursued in evolutionary dynamics in finite populations. Nevertheless, fixation trivially occurs in the presence of zealots if all zealots are assumed to take the same strategy; the zealots' strategy always fixates. For example, if there is a single zealous cooperator in the population, cooperation always fixates even in the conventional prisoner's dilemma game. However, in this adverse case, fixation of cooperation is expected to take long time; the relevant question here
is the fixation time \cite{Antal2006BMB,Traulsen2007JTheorBiol,Altrock2009NewJPhys,Altrock2010PhysRevE,Assaf2010JStatMech,Ewens2010book,WuAltrock2010PhysRevE,Altrock2011PlosComputBiol,Assaf2012PhysRevLett,Kreindler2013GamesEconBehav}. Here we examine the mean fixation time of the strategy selected by the zealots.
This quantity serves as a probe to understand the extent to which zealots influence non-zealous players in the population. The fixation time would be
affected by the payoff matrix, population size,
number of zealous players, and strength of selection.
We derive the asymptotic dependence of the mean fixation time on the population size when the fraction of zealots in the population is fixed.
Mathematically, we extend the approach
taken in \cite{Antal2006BMB} to the case with zealots.

\section{Model}

We assume a well-mixed population of $N+M$ players
under evolutionary dynamics defined as follows.
In each discrete time unit, each player selects
either of the two strategies $A$ or $B$. 
Each player plays
a symmetric two-person game with
all the other $N+M-1$
players in a unit time.
The payoff matrix of the single game
for the row player is given by
\begin{equation}
\bordermatrix{
 & A & B \cr
A & a & b \cr
B & c & d \cr}.
\label{eq:payoff matrix}
\end{equation}
The fitness of a player on which the selection pressure
operates is defined as the payoff summed over
the $N+M-1$ opponents.

We assume that $N$ players may flip the strategy according to the Moran process \cite{Moran1958PCPS,Ewens2010book}. We call these players the ordinary players. The other $M$ players are zealots that never change the strategy irrespectively of their fitness. Because our primary interest is in the possibility of cooperation in social dilemma games induced by zealous cooperators, we assume that all zealots take strategy $A$; $A$ is identified with cooperation in the case of a social dilemma game.
We also assume that $a, b, c, d \geq 0$ for the Moran process to be well-defined.

Because we have assumed a
well-mixed population, the state of the evolutionary process
is specified by the
number of ordinary players selecting $A$, which we
denote by $i$.  In each time step, we select an ordinary player with the equal probability $1/N$. The strategy of the selected player is updated.
Then, we select a player, called the parent, whose strategy replaces that of the previously selected player.
The parent is selected with the probability
proportional to the fitness among the $N+M$ players including the zealots 
and the player whose strategy is to be replaced.
The population size $N$ is constant over time. 
It should be noted that a player
is updated once on average in time $N$. 

Because the zealots always select $A$, the Moran process ends up with
the unanimous population of $A$ players (we impose
$a>0$ for this to be true). In other words,
fixation of $A$ always occurs such that the issue of
fixation probability is irrelevant to our model.
%
% Therefore, the primary question regarding our model is the fixation time.

\section{Results}

We calculate the mean fixation time and its approximation in the case of a large population size by extending the framework developed in \cite{Antal2006BMB} (also see \cite{Vankampen2007book,Redner2001book,Krapivsky2010book,Ewens2010book}).

\subsection{Mean fixation time: exact solution}

Consider the state of the population in which $i$ 
($0\le i\le N$) ordinary players select strategy $A$.
A total of $i+M$ and $N-i$ players, including the zealots, select strategies $A$ and $B$, respectively. The Moran process is equivalent to a random walk on the $i$ space in which $i=0$ is a reflecting boundary, and $i=N$ is the unique absorbing boundary.

The fitness of an $A$ and $B$ player is given by
\begin{equation}
f_i = \frac{(i+M-1)a + (N-i)b}{N+M-1}
\label{eq:fi}
\end{equation}
and
\begin{equation}
g_i = \frac{(i+M)c + (N-i-1)d}{N+M-1},
\label{eq:gi}
\end{equation}
respectively. In a single time step,
$i$ increases by one, does not change, or decreases by one.
We denote by $T^+_i$ and $T^-_i$ the probabilities that $i$ shifts to $i+1$ and $i-1$, respectively. These probabilities are given by
\begin{equation}
 T_i^+ = \frac{N-i}{N} \frac{(i+M)f_i}{(i+M)f_i + (N-i)g_i}
\label{eq:T^+_i}
\end{equation}
and
\begin{equation}
 T_i^- = \frac{i}{N} \frac{(N-i)g_i}{(i+M)f_i + (N-i)g_i}.
\label{eq:T^-_i}
\end{equation}

We denote by $t_i$ the mean fixation time when there are initially $i$ ordinary players with strategy $A$. As shown in \cite[pp. 86-91]{Ewens2010book} (see Appendix~\ref{sec:derivation of t_i} for a full derivation), we obtain
\begin{equation}
 t_i = \sum_{j=i}^{N-1} q_j\sum_{k=0}^j \frac{1}{T_k^+ q_k},
\label{eq:t_i}
\end{equation}
where
\begin{equation}
q_k = \prod^{k}_{j=1}\frac{T^{-}_{j}}{T^{+}_{j}}.
\label{eq:def q general}
\end{equation}
In Eq.~\eqref{eq:def q general}, we interpret $q_0 = 1$.

\subsection{Deterministic approximation of the random walk}\label{sub:deterministic}

In this section we classify
the deterministic dynamics driven by the expected bias of the random walk
(i.e., $T_i^+ - T_i^-$) into three cases, as is done in the analysis of populations without zealots \cite{Taylor2004BullMathBiol,Antal2006BMB}.
The obtained classification determines
the dependence of the mean fixation time on $N$, as we will show in section~\ref{sub:large N limit}.

We first identify the equilibrium points of the deterministic dynamics,
i.e., $i$ satisfying 
$T^+_i = T^-_i$. Equations~\eqref{eq:T^+_i} and \eqref{eq:T^-_i} indicate that $i=N$ always yields $T^+_i = T^-_i=0$, corresponding to the fact that $i=N$ is the unique absorbing state. Other equilibria are derived from
\begin{equation}
(i+M)\left[(i+M-1)a+(N-i)b\right]-i\left[(i+M)c+(N-i-1)d\right] = 0.
\label{eq:T_i^+ = T_i^-}
\end{equation}
We set
$y\equiv i/N$ ($0\le y< 1$), $m\equiv M/N$, and ignore $O(N^{-1})$ terms in
Eq.~\eqref{eq:T_i^+ = T_i^-} to obtain
\begin{equation}
f(y)\equiv (a-b-c+d)y^2 + \left[2ma + (1-m)b - mc -d \right]y +m^2 a + mb = 0.
\label{eq:f(y)}
\end{equation}

We define
\begin{align}
 \tilde{y} =& -\frac{1}{2(a-b-c+d)} \left[ 2ma + \left(1-m\right)b -mc -d \right],
\label{eq:tilde y}\\
 D =& \left[2ma+(1-m)b-mc-d\right]^2-4m(ma+b)(a-b-c+d),
\label{eq:D}\\
 y^\ast_1 =&
\begin{cases}
 \tilde{y} - \frac{\sqrt{D}}{2(a-b-c+d)} & (a-b-c+d\neq 0),\\
 -\frac{m(ma+b)}{2ma+(1-m)b-mc-d} & (a-b-c+d=0),
\end{cases}
\label{eq:y1}
\end{align}
and
\begin{align}
 y^\ast_2 =& \tilde{y} + \frac{\sqrt{D}}{2(a-b-c+d)}.
\label{eq:y2}
\end{align}
We will use $y^\ast_2$ only when $a-b-c+d>0$.
In the continuous state limit,
the deterministic dynamics driven by $T_i^+-T_i^-$ 
is classified into the following three cases, as summarized in Table~\ref{tab:cases 1-3}. The derivation is
shown in Appendix~\ref{sec:calc deterministic limits}.

\begin{figure}[htbp]
 \begin{center}
\includegraphics[width=12cm]{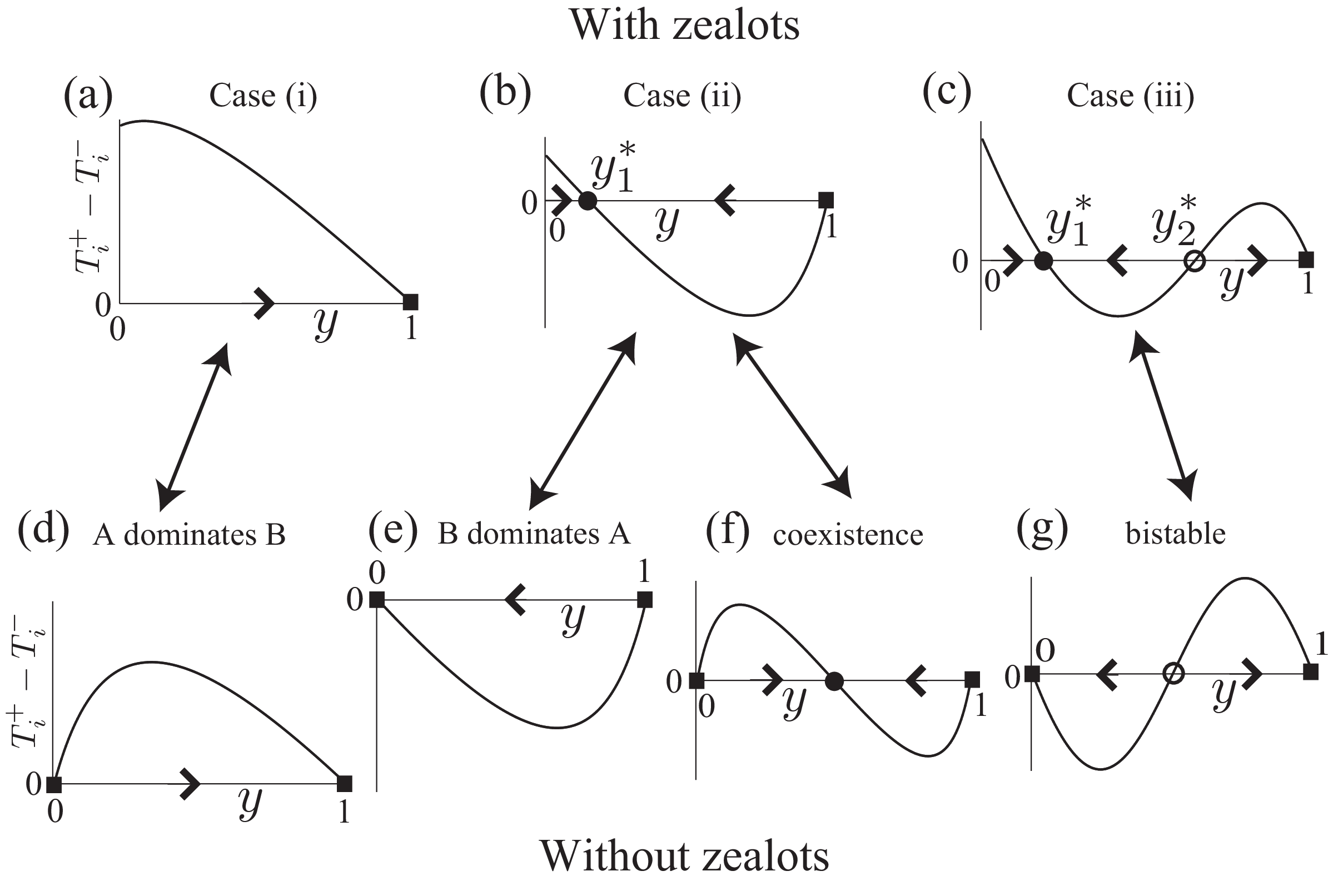}
 \end{center}
 \caption{Schematic classification of the deterministic dynamics driven by
$T_i^+-T_i^-$. (a)--(c) Populations with zealots. (d)--(g) Populations without zealots. Filled and open circles represent stable and unstable equilibria, respectively. Filled squares represent the absorbing boundary condition. It should be noted that we identify $y=i/N$.}
\label{fig:deterministic}
\end{figure}

\begin{enumerate}
\renewcommand{\labelenumi}{Case (\roman{enumi}):}
\item $f(y)> 0$ holds true for all $y$ ($0\le y\le 1$) such that the dynamics starting from any initial condition tends to $y=1$ (Fig.~\ref{fig:deterministic}(a)).
In an infinite population, $A$ dominates $B$.
In a finite population, we expect that the fixation time is short.
This case occurs when $c < (m+1)a$ and one of the following conditions is satisfied:
\begin{itemize}
\item $a-b-c+d \leq 0$.
\item $a-b-c+d > 0$ and $y_1^\ast\le 0$ (i.e., $2ma+(-m+1)b-mc-d\ge 0$).
\item $a-b-c+d > 0$, $0<y_1^\ast<1$ (i.e., $2ma+(-m+1)b-mc-d<0$ and $-(2m+2)a+(m+1)b+(m+2)c-d<0$), and $D \leq 0$.
\item $a-b-c+d > 0$ and $y_1^\ast\ge 1$ (i.e., $-(2m+2)a+(m+1)b+(m+2)c-d\ge 0$).
\end{itemize}

\item $f(y)=0$ has a unique solution $y^\ast_1$ ($0< y^\ast_1<1$) such that
the dynamics starting from any initial condition converges to $y^\ast_1$
(Fig.~\ref{fig:deterministic}(b)).
In an infinite population, $A$ and $B$ coexist.
In a finite population, we expect that the fixation time is long.
This case occurs when $c > (m+1)a$.

\item $f(y)=0$ has two solutions $0<y^\ast_1 < y^\ast_2<1$.
Dynamics starting from $0\le y<y^\ast_2$ converges to $y^\ast_1$, and
that starting from $y^\ast_2<y<1$ converges to $y=1$
(Fig.~\ref{fig:deterministic}(c)).
In an infinite population, a mixture of $A$ and $B$ and the pure $A$ configuration are bistable.
In a finite population, we expect that the fixation time is long if the dynamics starts with $0\le y<y^\ast_2$ and short if it starts with $y^\ast_2<y<1$.
This case occurs when
\begin{align}
c <& (m+1)a,\label{eq:c<(m+1)a}\\
a-b-c+d >& 0,\\
0<&\tilde{y}<1,\label{eq:0 < tilde y < 1}
\end{align}
and
\begin{equation}
D > 0
\label{eq:D>0}
\end{equation}
are satisfied.

\end{enumerate}

The condition given by Eq.~\eqref{eq:c<(m+1)a}
is related to the so-called cooperation facilitator assumed in a previous model
\cite{Mobilia2012PhysRevE} as follows. 
Consider a hypothetical infinite population in which
almost all players select $A$, i.e., $y\approx 1$.
Then, the payoff that a player with strategy $A$ gains by being matched with the other
ordinary players and zealous players is equal to
$(m+1)a$. The payoff that a player with strategy $B$ gains by being matched with the other ordinary players, but not zealous players, is equal to $c$. Therefore, Eq.~\eqref{eq:c<(m+1)a} represents the condition for the stability
of the homogeneous population of strategy $A$ against invasion by $B$
when zealous players somehow contribute to the payoff of ordinary $A$ players and not to that of ordinary $B$ players. Such a zealous player is equivalent to the cooperation facilitator assumed in Ref.~\cite{Mobilia2012PhysRevE}.

In the corresponding model without zealots,
there are four scenarios: $A$ dominates $B$ (Fig.~\ref{fig:deterministic}(d)),
$B$ dominates $A$ (Fig.~\ref{fig:deterministic}(e)),
a mixture of $A$ and $B$ is stable (Fig.~\ref{fig:deterministic}(f)),
and $A$ and $B$ are bistable (Fig.~\ref{fig:deterministic}(g))
\cite{Antal2006BMB}.
The cases shown in Fig.~\ref{fig:deterministic}(d),
Fig.~\ref{fig:deterministic}(f), and
Fig.~\ref{fig:deterministic}(g) are analogous to
cases (i), (ii), and (iii), respectively, for the game with zealots.
The case shown in Fig.~\ref{fig:deterministic}(e) never occurs
in the game with zealots because $y$ tends to increase
in the absence of $A$ owing to the fact that
unanimity of $B$ among the ordinary players is a reflecting boundary of our model. In fact, this case corresponds to case (ii) for the presence of zealots
(Fig.~\ref{fig:deterministic}(b)).
If we set $m\to 0$, we obtain case (i) when
$a-c>0$ and $b-d>0$,
case (ii) when $a-c<0$, and case (iii) when
$a-c>0$ and $b-d<0$. As is consistent with \cite{Antal2006BMB}, the
classification depends only on the $a-c$ and $b-d$ values. However, 
the scenario in which $B$ dominates
$A$ (Fig.~\ref{fig:deterministic}(e)) does not happen
even with the vanishing density of zealots (i.e., $m\to 0$) because
the unanimity of $B$ remains to be a reflecting boundary as long as there is at least one zealot.

\subsection{Mean fixation time: large $N$ limit}\label{sub:large N limit}

In this section, we analyze the order of the mean fixation time in terms of $N$ when $N$ is large. We assume that the fraction of zealots in the population, i.e., $m=M/N$, is fixed.
Because the mean fixation time is by definition the largest for $i=0$, i.e., the initial condition in which all ordinary players select $B$, we focus on $t_0$.
To evaluate $t_0$, we rewrite Eq.~\eqref{eq:t_i} for $i=0$ as
\begin{equation}
t_0 = \sum_{k=0}^{N-1} \frac{1}{T_k^+q_k}\sum_{j=k}^{N-1}q_j.
\label{eq:t_0}
\end{equation}

\subsubsection{Case (i)} \label{thecase1}

We obtain
\begin{equation}
\frac{T_i^-}{T_i^+} = 1- \frac{Nf(i/N)}{(i+M)[(i+M-1)a+(N-i)b]},
\end{equation}
where $f(y)$ ($0\le y<1$) is given by Eq.~\eqref{eq:f(y)}.
In case (i), $f(y)>0$ holds true. Therefore, 
\begin{align}
\frac{T_i^-}{T_i^+} \le& \sup_{0\le y<1} 
\left\{1- \frac{f(y)}{(y+m)[(y+m)a+(1-y)b]}\right\}\notag\\
\equiv& \varepsilon < 1
\label{eq:def epsilon}
\end{align}
is satisfied for $0\le i\le N-1$. By using Eq.~\eqref{eq:def epsilon}, we obtain
\begin{align}
\frac{1}{q_k}\sum_{j=k}^{N-1}q_j =& \sum_{j=k}^{N-1}\prod_{\ell=k+1}^j\frac{T_\ell^-}{T_\ell^+}\notag\\
\leq& \sum_{j=k}^{N-1} \varepsilon^{(j-k-1)}
%
% =& \frac{1}{\varepsilon}\frac{1}{1-\varepsilon}(1-\varepsilon^{N-k}) \\
%
\leq \frac{1}{\varepsilon}\frac{1}{1-\varepsilon}.
\label{eq:bound case (i)}
\end{align}
Because the left-hand side of Eq.~\eqref{eq:bound case (i)} is at least unity, we obtain
\begin{equation}
 t_0 \propto \sum_{k=0}^{N-1} \frac{1}{T_k^+}.
\end{equation}
The substitution of $y = i/N$ and $m = M/N$ in Eq.~\eqref{eq:T^+_i} yields
\begin{equation}
\frac{1}{T_i^+} = \frac{1}{1-y} + \frac{(y+m)c+(1-y)d}{(y+m)\left[
(y+m)a+(1-y)b\right]}.
\label{eq:1/T_i^+}
\end{equation}
In particular, we obtain
\begin{equation}
\frac{1}{T^+_i} \approx \frac{1}{1-y}\quad (y \approx 1).
\label{eq:singularity of 1/T_i^+}
\end{equation}
Equation~\eqref{eq:singularity of 1/T_i^+} implies that
\begin{equation}
t_0 \propto N\ln N.
\label{eq:t_0 order}
\end{equation}
This result coincides with the previous result for the absence of zealots \cite{Antal2006BMB}.

\subsubsection{Case (ii)}\label{thecase2}

In Case (ii), $T^+_i-T^-_i>0$ for $0\le i<N y^\ast_1$ and
$T^+_i-T^-_i <0$ for $N y^\ast_1 <i<N$.
Therefore, $q_i$ takes the minimum at $i\approx N y^\ast_1$.
We denote the value of $i$ that satisfies $i<N y^\ast_1$ and $q_i\approx q_{N-1}$
by $i^\ast$. Such an $i^\ast$ exists if $q_0\ge q_{N-1}$. If $q_0<q_{N-1}$, we regard that $i^\ast=0$. Using the relationship 
$q_i = \left[\tilde{q}(i/N)\right]^N$ for a function $\tilde{q}(y)$ ($0\le y<1$)
\cite{Antal2006BMB} (also see Appendix~\ref{sec:tilde q}), 
we obtain
\begin{align}
t_0 =& \sum_{k=0}^{N-1} \frac{1}{T^+_k q_k}\sum_{j=k}^{N-1}q_j\notag\\
\sim& \sum_{k=0}^{N-1} \frac{1}{T^+_k q_k} \max\{q_k, q_{N-1}\}\notag\\
\sim& \sum_{k=i^\ast}^{N-1} \frac{q_{N-1}}{q_k}\notag\\
\sim& \sqrt{N} \left[\frac{\tilde{q}(1)}{\tilde{q}(y^\ast)} \right]^N,
\label{eq:t_0 exponential}
\end{align}
where
\begin{equation}
\tilde{q}(y^\ast)=\min_{0\le y<1}\tilde{q}(y).
\label{eq:def tilde q y*}
\end{equation}
To derive the last line in Eq.~\eqref{eq:t_0 exponential},
we used the steepest descent method \cite{Antal2006BMB} (also see Appendix~\ref{sec:steepest descent method}). 

\begin{figure}[htbp]
 \begin{center}
\includegraphics[width=6cm]{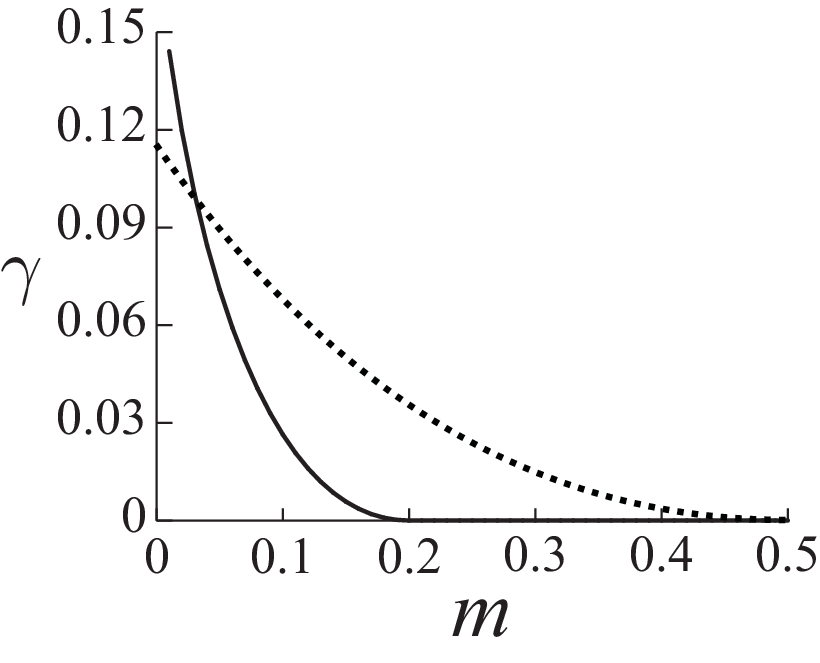}
 \end{center}
 \caption{The exponent $\gamma$ for the mean fixation time (Eq.~\eqref{eq:t_0 case (ii)}) plotted against the density of zealots $m$ for the prisoner's dilemma game with $a=1$, $b=0$, $c=1.2$, and $d=0$
(solid line) and the snowdrift game with
$a=\beta-0.5$, $b=\beta$, $c=\beta-1$, $d=0$, with $\beta=1.5$ (dotted line). We calculated $\gamma$ on the basis of Eqs.~\eqref{eq:t_0 exponential}, \eqref{eq:def tilde q y*}, and \eqref{eq:q_i rewrite}.}
\label{fig:gamma vs m}
\end{figure}

Equations~\eqref{eq:1/T_i^+} and \eqref{eq:singularity of 1/T_i^+} imply
that $1/T_k^+$ in Eq.~\eqref{eq:t_0 exponential} is safely ignored near the singularity at $y\approx 1$ because it would contribute at most $\propto N\ln N$ to the fixation time. Therefore, we obtain
\begin{equation}
t_0 \propto \sqrt{N}\exp(\gamma N),
\label{eq:t_0 case (ii)}
\end{equation}
where $\gamma > 0$ is a constant that depends on
$a, b, c, d$, and $m$.
The dependence of $\gamma$ on $m$ is shown in Fig.~\ref{fig:gamma vs m} for sample payoff matrices for the prisoner's dilemma game (solid line)
and snowdrift game (dotted line). For both games, $\gamma$ monotonically decreases with $m$, implying that the fixation time decreases with $m$. In particular, $\gamma$ is equal to zero, which corresponds to
$t_0 \propto N\ln N$, when $m$ is larger than a threshold value.

\subsubsection{Case (iii)}

In this case, $q_i$ takes a local minimum at $i=Ny^\ast_1$ and
a local maximum at $i=Ny^\ast_2$.
Therefore, behavior of the random walk in the range
$0\le i <Ny^\ast_2$ is qualitatively the same as that for case (ii),
and that in the range $Ny^\ast_2<i<N$
is qualitatively the same as that for case (i).
Because the former part makes the dominant contribution to the 
fixation time, the scaling of the mean fixation time is given by
Eq.~\eqref{eq:t_0 case (ii)}.

Case (iii) occurs when strategy $A$ is disadvantageous when it is rare and advantageous when it is frequent. The coordination game provides such an example (section~\ref{sub:coordination game}).

\subsubsection{Summary and the borderline case}

In summary, the mean fixation time in the limit of large $N$ is given by 
$t_0 \propto N\ln N$ in case (i) and $t_0 \propto \sqrt{N}\exp(\gamma N)$ ($\gamma>0$) in cases (ii) and (iii).
For the parameter values on the boundary between the two scaling regimes,
the same arguments as those for the model without zealots \cite{Antal2006BMB}
lead to $t_0 \propto N^{3/2}$.

\section{Dependence of the mean fixation time on the selection strength}

We examine the influence of the selection strength, denoted by $w$, on the
mean fixation time. To this end,
we redefine the fitness to an $A$ and $B$ player by
$1-w+wf_i$ and $1-w+wg_i$, respectively, where $f_i$ and $g_i$ are given by
Eqs.~\eqref{eq:fi} and \eqref{eq:gi} (e.g., \cite{Nowak2004Nature,Nowak2006book}).
Consequently, we replace the payoff matrix given by
Eq.~\eqref{eq:payoff matrix} by
\begin{equation}
\bordermatrix{
 & A & B \cr
A & 1-w+wa & 1-w+wb \cr
B & 1-w+wc & 1-w+wd \cr}.
\label{eq:payoff matrix w}
\end{equation}
Equation~\eqref{eq:payoff matrix} is reproduced with $w=1$.

For sufficiently weak selection, we obtain $t_0 \propto N\ln N$, i.e., case (i), regardless of the payoff matrix. To prove this statement,
we note that, by using the payoff matrix shown in Eq.~\eqref{eq:payoff matrix w}, condition $c < (m+1)a$ in the case of $w=1$ is generalized to
\begin{equation}
c < (m+1)a+m\left(\frac{1}{w}-1\right).
\label{eq:c'a'}
\end{equation}
Therefore, if the original game in the case of $w=1$ belongs case (ii),
i.e., $c > (m+1)a$, the game belongs to case (i) or (iii)
(Table~\ref{tab:cases 1-3}) if Eq.~\eqref{eq:c'a'}, or equivalently,
\begin{equation}
w < w_1\equiv \frac{m}{c-(m+1)a+m}
\label{eq:w_c 1}
\end{equation}
is satisfied. For a fixed payoff matrix, $w_1$ monotonically increases with $m$, consistent with the intuition that existence of
zealots would lessen the fixation time.

Next, 
the sign of $a-b-c+d$ is not affected by the selection strength. Therefore,
we assume $a-b-c+d>0$ and prove that a condition for
case (iii), i.e., Eq.~\eqref{eq:D>0}, is violated with a sufficiently small $w$.
Because the value of $\tilde{y}$ given by
Eq.~\eqref{eq:tilde y} is also unaffected by $w$, we start with assuming
$0<\tilde{y}<1$, which is a necessary condition for case (iii)
(Eq.~\eqref{eq:0 < tilde y < 1}; see Appendix~\ref{sec:calc deterministic limits}).
The condition $D< 0$ in the case of $w=1$, where $D$ is defined by Eq.~\eqref{eq:D}, is generalized to
\begin{equation}
w D -4 (1-w)m(m+1)(a-b-c+d)< 0.
\label{eq:D'}
\end{equation}
Because the condition imposed on $D$, which distinguishes cases (i) and (iii),
is relevant only for
$a-b-c+d>0$ (Table~\ref{tab:cases 1-3}), Eq.~\eqref{eq:D'} is satisfied for an arbitrary payoff matrix if
\begin{equation}
w < w_2\equiv \frac{4m(m+1)(a-b-c+d)}{D+4m(m+1)(a-b-c+d)}.
\label{eq:w_c 2}
\end{equation}
Therefore, case (iii) is excluded with a sufficiently small $w$ value.
% \blue{For a fixed payoff matrix, $w_2$ monotonically increases with $m$.} \naoki{True?}

The threshold value of $w$ below which $t_0 \propto N\ln N$, which we denote by
$w_{\rm c}$, is given by
\begin{equation}
w_{\rm c}=\begin{cases}
\min\{w_1, w_2, 1\} & (w_1>0, w_2>0),\\
\min\{w_1, 1\} & (w_1>0, w_2<0),\\
\min\{w_2, 1\} & (w_1<0, w_2>0),\\
1 & (w_1<0, w_2<0). 
\end{cases}
\label{eq:w_c}
\end{equation}

We can alternatively introduce the selection strength
by replacing Eqs.~\eqref{eq:fi} and \eqref{eq:gi} to redefine the fitness by
\begin{equation}
f_i = \exp\left[\beta \frac{(i+M-1)a + (N-i)b}{N+M-1}\right]
\label{eq:fi beta}
\end{equation}
and
\begin{equation}
g_i = \exp\left[\beta \frac{(i+M)c + (N-i-1)d}{N+M-1}\right],
\label{eq:gi beta}
\end{equation}
where $\beta$ is the selection strength \cite{Traulsen2008BullMathBiol}.
In Appendix~\ref{sec:proof with beta}, we show that qualitatively the same result holds true in the sense that
there is a threshold value of $\beta$ below which the fixation is fast irrespective of the $a$, $b$, $c$, $d$, and $m$ values. It should be noted that, with Eqs.~\eqref{eq:fi beta} and \eqref{eq:gi beta}, $a$, $b$, $c$, and $d$ are allowed to take negative values.

\section{Examples}

We compare the mean fixation time for some games with that for the
neutral game, i.e., $a = b = c = d >0$. In the absence of zealots,
the neural game yields $T_i^+=T_i^-$ 
($1\le i\le N-1$).
The random walk is unbiased, and
the so-called mean conditional fixation time is equal to $N(N-1)$ \cite{Antal2006BMB}. The mean conditional fixation time is defined as the mean fixation time starting from state $i=1$ under the condition that the absorbing state at $i=N$, not $i=0$, is reached.

The neutral game in the presence of zealots yields $T_0^+ > T_0^- = 0$ and
$T_i^+/T_i^-=(i+M)/i$ ($1\le i\le N-1$).
Therefore, the random walk is biased toward $i=N$ for all $i$.
More precisely, we obtain
\begin{equation}
t_0 = N(N+M)\sum_{0\leq k\leq i\leq N-1} \frac{i!(k+M)!}{k!(i+M)!}\frac{1}{(N-i)(i+M)}.
\label{eq:t_0 neutral game}
\end{equation}
As in Ref.~\cite{Antal2006BMB}, we say that fixation is fast (slow)
if $t_0$ is smaller (larger) than the value given by 
Eq.~\eqref{eq:t_0 neutral game}.
It should be noted that
$t_0 \propto N\ln N$ for the neutral game because it corresponds to $w=0<w_{\rm c}$. 

\subsection{Constant selection}

As a first example, consider the case of frequency-independent selection such that $A$ and $B$ are equipped with fitness $r$ and 1 (under $w=1$), respectively. 
When $a=b=r$ and $c=d=1$,
the threshold selection strength below which
$t_0 \propto N\ln N$, i.e., case (i), holds true is given by
\begin{equation}
w_{\rm c} = \begin{cases}
\frac{m}{(m+1)(1-r)} & \left(r\le \frac{1}{m+1}\right),\\
1 & \left(r\ge \frac{1}{m+1}\right). 
\end{cases}
\label{eq:w_c for constant selection}
\end{equation}
If $w>w_{\rm c}$, case (ii) occurs.
Even if $A$ is disadvantageous to $B$, $A$ fixates fast with the help of zealots regardless of the selection strength if $1/(m+1)<r<1$.
This condition is more easily satisfied when $m$ is larger.

\subsection{Prisoner's dilemma game}

Consider the prisoner's dilemma game with a standard payoff matrix given by
$a=1$, $b=0$, $c=T$, and $d=0$, where $T>1$. Strategies $A$ and $B$ represent cooperation and defection, respectively. 
It should be noted that $a-b-c+d < 0$.
With a general selection strength, the conditions derived in
section~\ref{sub:deterministic} imply that
$t_0 \propto N\ln N$, i.e., case (i), if $T < 1+m/w$, and
$t_0 \propto \sqrt{N}\exp(\gamma N)$ with case (ii) if $T > 1+m/w$.
This condition coincides with that for the dominance of cooperators in the case of the infinite population \cite{Masuda2012SciRep}.

The mean fixation time with $w=1$ and $m=0.2$ obtained by direct calculations of Eq.~\eqref{eq:t_0} is shown in 
Fig.~\ref{fig:t_0 PD}(a). In this and the following figures, the $t_0$ values are those normalized by that for the neutral game [Eq.~\eqref{eq:t_0 neutral game}]. The behavior of $t_0$ is qualitatively different according to whether $T$ is larger or smaller than $1+m/w=1.2$. If $T<1.2$, the ratio of $t_0$ for the prisoner's dilemma game to $t_0$ for the neutral game seems to approach a constant as $N\to\infty$. This is consistent with case (i). In contrast, if $T>1.2$, $t_0$ grows rapidly, which is consistent with case (ii).
To be more quantitative, $400 \sqrt{N}\exp(\gamma N)$ divided by the $t_0$ value for the neutral game is shown by the dashed line in Fig.~\ref{fig:t_0 PD}(a). It should be noted that 400 is a constant for fitting and that $\gamma$ value is theoretically determined as described in section~\ref{thecase2}. The theory (dashed line) agrees well with the exact numerical results (thinnest solid line).
We remark that the normalized $t_0$ behaves non-monotonically in $N$; it takes a minimum at an intermediate value of $N$.

Next, to examine the effect of the selection strength,
we set $T=1.2$ and $m = 0.1$.
The mean fixation time as a function of $N$ and $w$ is shown in
Fig.~\ref{fig:t_0 PD}(b).
Equation~\eqref{eq:w_c} implies that
$t_0 \propto N\ln N$ when $w < w_{\rm c}= 0.5$. Consistent with this result,
$t_0$ grows fast as a function of $N$ when $w$ is large (i.e., $w=0.7$ and 1).
In particular, for $w=1$,
$400 \sqrt{N}\exp(\gamma N)$ normalized by the $t_0$ value for the neutral game
(dashed line in Fig.~\ref{fig:t_0 PD}(b)) agrees well with the exact results
(thin solid line).
For small $w$ (i.e., $w=0.4$), $t_0$ seems to scale with $N\ln N$ (thick solid line).

Figure~\ref{fig:t_0 PD}(c) shows the dependence of $t_0$ on $N$ for different
densities of zealots (i.e., $m$). It should be noted that the baseline $t_0$ value derived from the neutral game depends on the value of $m$. Because we set $T=1.2$ and $w=1$ in Fig.~\ref{fig:t_0 PD}(c), the threshold value of $m$ is equal to 0.2. In fact, the normalized $t_0$ diverges according to
$\propto \sqrt{N}\exp(\gamma N)$ when $m=0.1$ (dashed line and thick solid line), whereas it seems to converge to a constant value when $m=0.3$ (thin solid line).

Figure~\ref{fig:t_0 PD} indicates that $t_0$ for the prisoner's dilemma game is always larger than that for the neutral game (i.e., the normalized $t_0$ is larger than unity). This is consistent with the intuition that cooperation is difficult to attain in the prisoner's dilemma game as compared to the neutral game.

Finally, consider the symmetrized donation game, which is another standard form of the prisoner's dilemma game, given by $a=b^{\prime}-c^{\prime}$, $b=-c^{\prime}$,
$c=b^{\prime}$, and $d=0$, where $b^{\prime}$ is the benefit, and $c^{\prime} (<b^{\prime})$ is the cost. For the Moran process to be well-defined, we require
$1-w+wb\ge 0$, i.e., $w<1/(1+c^{\prime})$.
For this payoff matrix, we obtain
\begin{equation}
w_{\rm c}=\begin{cases}
\frac{m}{m-b^{\prime}+(1+m)c^{\prime}} & \left(\frac{b^{\prime}}{c^{\prime}}\le \frac{m+1}{m}\right),\\
1 & \left(\frac{b^{\prime}}{c^{\prime}}\ge \frac{m+1}{m}\right).
\end{cases}
\end{equation}
Fixation occurs fast for a large benefit-to-cost ratio, large $m$, or small selection strength.

\subsection{Snowdrift game}

In this section, we examine the snowdrift game \cite{Maynardsmith1982book,Sugden1986book,Hauert2004Nature} defined by
$a=\beta-0.5$, $b=\beta-1$, $c=\beta$, and $d=0$, where $\beta>1$. Strategies $A$ an $B$ are identified as cooperation and defection, respectively. Each player is tempted to defect if the other player cooperates, as in the prisoner's dilemma game. However, different from the prisoner's dilemma game, a player is better off by cooperating if the partner defects; mutual defection is the worst outcome. In the infinite well-mixed population without zealots, the game has the unique 
mixed Nash equilibrium in which the fraction of cooperation is equal to $(2\beta-2)/(2\beta-1)$. 

Numerical evidence for the replicator dynamics, corresponding to an infinite population, suggests that cooperation is dominant
if $m$ is large or $w$ is small \cite{Masuda2012SciRep}.
For the finite population, we obtain
\begin{equation}
w_{\rm c} = \begin{cases}
\frac{2m}{3m-2m\beta+1} & \left(\beta\le \frac{m+1}{2m}\right),\\
1 & \left(\beta\ge \frac{m+1}{2m}\right).
\end{cases}
\end{equation}
If $w<w_{\rm c}$, we obtain $t_0 \propto N\ln N$, i.e., case (i).
If $w>w_{\rm c}$, we obtain $t_0 \propto \sqrt{N}\exp(\gamma N)$ with case (ii).
A large value of $\beta$ or $m$ makes the fixation time smaller. This result makes sense because a large $\beta$ generally favors cooperation.

\subsection{Coordination game}\label{sub:coordination game}

The coordination game given by $a=d>0$ and $b=c=0$ has two pure 
Nash equilibria in the infinite well-mixed population without zealots. For a finite population in the presence of zealots, Eq.~\eqref{eq:w_c} yields
\begin{equation}
w_{\rm c} = \begin{cases}
\frac{8m(m+1)}{a(-4m^2-4m+1)+8m(m+1)} & \left(0\le m\le \frac{\sqrt{2}-1}{2}\right),\\
1 & \left(\frac{\sqrt{2}-1}{2}\le m\le 1\right).
\end{cases}
\label{eq:w_c coordination game}
\end{equation}
If $w<w_{\rm c}$, we obtain $t_0 \propto N\ln N$, i.e., case (i).
It should be noted that any strength of selection
$0\le w\le 1$ yields $t_0 \propto N\ln N$ if there are sufficiently many zealots, similar to the game with constant selection, prisoner's dilemma game, and snowdrift game.
If $w>w_{\rm c}$, we obtain $t_0 \propto \sqrt{N}\exp(\gamma N)$ with case (iii).

\begin{figure}
 \begin{center}
  \includegraphics[width=8cm]{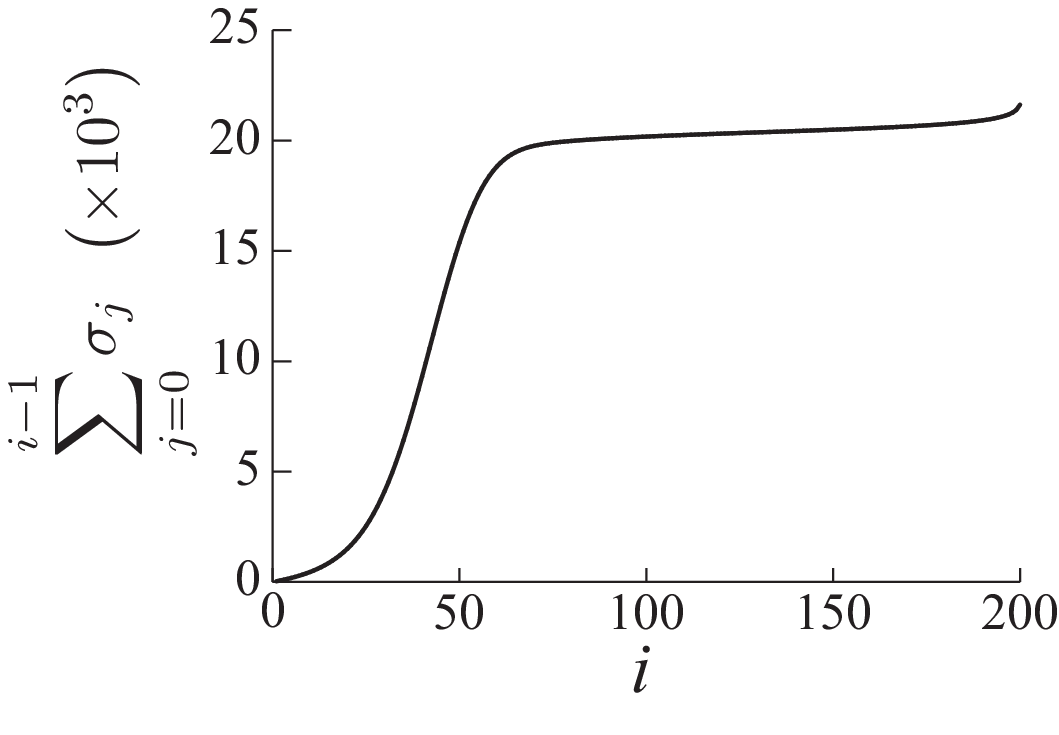}
 \end{center}
 \caption{Mean first-passage time for the coordination game. We set
$N=200$, $a=d=1$, $b=c=0$, $m=0.2$, and $w=1$.}
\label{fig:coordination game}
\end{figure}

The mean first-passage time from state 0 (i.e., all ordinary players select
$B$) to state $i$, i.e., $\sum_{j=0}^{i-1} \sigma_j$, is shown in
Fig.~\ref{fig:coordination game}. 
It should be noted that $t_0$ is equal to this first-passage time to exit
$i=N$. We set $N=200$, $a=d=1$, $b=c=0$, $m=0.2$, and $w=1$.
Equation~\eqref{eq:w_c coordination game} implies
$w_{\rm c}=48/49$ for these parameter values. Because $w=1>w_{\rm c}$, we 
obtain case (iii).

The first-passage time increases slowly
as $i$ increases when $i$ is small.
It rapidly increases with $i$ for intermediate values of $i$,
Once the random walker passes the critical $i$ value, it feels a positive bias such that the first-passage time only gradually increases with $i$ for large $i$.
The values of $i$ that separate the three regimes are roughly consistent with
the analytical estimates $y^\ast_1 = 0.1$ and $y^\ast_2 = 0.2$ [Eqs.~\eqref{eq:y1} and \eqref{eq:y2}].
It should be noted that the first-passage time shows representative behavior of case (iii) although $w$ is only slightly larger than $w_{\rm c}$.

\section{Discussion}

We extended the results for the fixation time under the Moran process \cite{Antal2006BMB} to the case of a population with zealous players.
Similar to the case without zealots \cite{Antal2006BMB}, we identified three regimes in terms of the payoff matrix, number of zealots, and selection strengths. In one regime, the fixation time is small (i.e., $\propto N\ln N$). In the other two regimes, it is large (i.e., $\propto \sqrt{N}\exp(\gamma N)$ with $\gamma>0$). We illustrated our results with representative games including the prisoner's dilemma game, snowdrift game, and coordination game. 

Zealots have several impacts on evolutionary dynamics in finite populations. First, fixation of one strategy $A$ always occurs with zealots because we assumed that all zealots permanently take $A$. 
Second, there is a case in which fixation is fast if the fraction of $A$ players is sufficiently large, whereas fixation is slow if the fraction of $A$ is small. This scenario occurs for the coordination game. In the absence of zealots, the same game shows bistability such that the fixation to the unanimity of $A$ or that of $B$ occurs fast \cite{Antal2006BMB}.
Third, for a selection strength smaller than a threshold value, the fixation is fast for any payoff matrix.
In the absence of zealots, the dependence of the mean fixation time on $N$ for large $N$ values is completely determined by the signs of $a-c$ and $b-d$ \cite{Antal2006BMB}. Therefore, the scaling of the mean fixation time on $N$
is independent of the selection strength because manipulating the selection strength does not change the sign of the effective $a-c$ or $b-d$ value. If the payoff matrix is given in the slow fixation regime, the fixation is exponentially slow even for a small selection strength. In contrast, in the presence of zealots, slow fixation can be accelerated if we lessen the selection strength.

Mobilia examined the prisoner's dilemma game with cooperation facilitators
\cite{Mobilia2012PhysRevE}. A cooperation facilitator was assumed to cooperate with cooperators and not to play with defectors.
The cooperation facilitator and zealous cooperator
 in the present study are common 
in that they never change the strategy. However, they
are different. First, zealous cooperators are embedded in a well-mixed population such that they myopically cooperate with defectors as well as cooperators. 
Second, the ordinary players may imitate the zealous cooperator's strategy (i.e., cooperation). In contrast, players do not imitate the cooperation facilitator's strategy (i.e., cooperation) in Mobilia's model.
As a consequence, cooperation does not always fixate in his model.

Examination of the case of imperfect zealots, in which zealots change the strategy with a small probability \cite{Masuda2012SciRep}, warrants future work.

\newpage

\setcounter{section}{0}
\renewcommand{\thesection}{\Alph{section}}

\newcounter{appendixcount}
\renewcommand{\theappendixcount}{\Alph{appendixcount}}
\newcommand{\appendixcount}{\refstepcounter{appendixcount}}

\section*{Appendix A: Derivation of Eq.~\eqref{eq:t_i}}\appendixcount\label{sec:derivation of t_i}

Denote by $P_i(t)$ the probability that the random walker
starting from state $i$ at time 0 is absorbed to state $N$ at time $t$. The normalization is given by $\sum^{\infty}_{t=0} P_i(t) = 1$.
It should be noted that $P_N(0)=1$ and $P_N(t)=0$ ($t\ge 1$).
The mean fixation time when $i$ ordinary players initially select strategy $A$
is given by
\begin{equation}
t_i = \sum^{\infty}_{t=0}tP_i(t).
\end{equation}
It should be noted that $t_N = 0$.

$P_i(t)$ satisfies the recursion relation given by
\begin{equation}
P_i(t) = T_i^- P_{i-1}(t-1) + (1-T_i^--T_i^+)P_i(t-1)+T_i^+ P_{i+1}(t-1).
\label{eq:recursion P}
\end{equation}
By multiplying both sides of Eq.~\eqref{eq:recursion P} by $t$ and taking the summation over $t$, we obtain
\begin{equation}
t_i = T_i^- t_{i-1} + (1-T_i^- - T_i^+)t_i+T_i^+ t_{i+1}+1.
\label{eq:recursion t}
\end{equation}
In terms of $\sigma_i \equiv t_i-t_{i+1}$, Eq.~\eqref{eq:recursion t} can be rewritten as
\begin{equation}
T^-_i\sigma_{i-1} - T^+_i\sigma_i + 1 = 0.
\label{eq:recursion sigma}
\end{equation}
The solution of Eq.~\eqref{eq:recursion sigma} is given by
\begin{equation} 
\sigma_i = \sigma_0 q_i + q_i\sum_{k=1}^i \frac{1}{T^+_kq_k},
\label{eq:sigma_i recursion}
\end{equation}
where $0\le i\le N-1$ and $q_i$ is given by Eq.~\eqref{eq:def q general}.

We set $i=0$ in Eq.~\eqref{eq:recursion t} and
use $T_0^- = 0$ to obtain
\begin{equation}
t_0 = (1-T_0^+)t_0+T_0^+ t_1+1.
\end{equation}
Therefore, we obtain
\begin{equation}
\sigma_0 = t_0-t_1 = \frac{1}{T_0^+}.
\label{eq:sigma_0}
\end{equation}
Using Eq.~\eqref{eq:sigma_0}, we reduce
Eq.~\eqref{eq:sigma_i recursion} to
\begin{equation}
\sigma_i = q_i\sum_{k=0}^i \frac{1}{T_k^+ q_k}.
\end{equation}
The mean fixation time is given by
\begin{align}
 t_i =& \sum_{j=i}^{N-1}\sigma_j + t_N\notag\\
=& \sum_{j=i}^{N-1} q_j\sum_{k=0}^j \frac{1}{T_k^+ q_k}.
\label{eq:t_i appendix}
\end{align}

\section*{Appendix B: Classification of the deterministic dynamics induced by the biased random walk}\appendixcount\label{sec:calc deterministic limits}

\renewcommand{\thesubsection}{B.\arabic{subsection}}

\subsection{When $a-b-c+d < 0$}

We obtain ${\rm d}^2f(y)/{\rm d}y^2<0$
for $a-b-c+d < 0$. Because
\begin{align}
f(0) =& m^2 a + mb > 0,\label{eq:f(0)>0}\\
f(1) =& (m+1)^2 a - (m+1)c,\label{eq:f(1)}
\end{align}
where we used the assumption $a>0$
in Eq.~\eqref{eq:f(0)>0},
we distinguish the following two cases.
If $c < (m+1)a$, $f(y)> 0$ is satisfied for $0\le y\le 1$, yielding case (i) in the main text. If $c > (m+1)a$, 
a certain $y^\ast_1 (0<y^\ast_1<1)$ exists such that
$f(y)>0$ for $0\le y<y^\ast_1$, and $f(y)<0$ for $y^\ast_1 < y\le 1$. Therefore, case (ii) occurs.

\subsection{When $a-b-c+d > 0$}

We obtain ${\rm d}^2f(y)/{\rm d}y^2>0$ for
$a-b-c+d > 0$. In this situation,
Eq.~\eqref{eq:f(0)>0} holds true.

If $f(1) < 0$, i.e., $c > (m+1)a$, a certain $y^\ast_1$ ($0<y^\ast_1<1$)
exists such that $f(y)>0$ for $0\le y<y^\ast_1$, and $f(y)<0$ for $y^\ast_1 < y\le 1$. Therefore, case (ii) occurs.

Suppose that $f(1) > 0$, i.e., $c<(m+1)a$.
To analyze this case, let us write
\begin{equation}
f(y) = (a-b-c+d)(y-\tilde{y})^2 + m^2 a + mb -\frac{\left[2ma+(1-m)b-mc-d\right]^2}{4(a-b-c+d)},
\end{equation}
where
\begin{equation}
\tilde{y} = -\frac{2ma + (1-m)b - mc - d}{2(a-b-c+d)}.
\end{equation}

\begin{enumerate}
\renewcommand{\labelenumi}{(\roman{enumi})}
\item If $\tilde{y} \leq 0$, i.e., $2ma+(-m+1)b-mc-d\ge 0$, we obtain $f(y) \geq f(0) > 0$ for $y\ge 0$. Therefore, case (i) occurs.
\item If $\tilde{y} \geq 1$, i.e., $-(2m+2)a+(m+1)b+(m+2)c-d\ge 0$, then $f(y) \geq f(1) > 0$, yielding case (i).
\item If $0 < \tilde{y} < 1$, we have the following two subcases:
\begin{enumerate}
\renewcommand{\labelenumi}{(\roman{enumi})}
\item If $D = \left[2ma+(1-m)b-mc-d\right]^2 - 4m(ma+b)(a-b-c+d) >0$,
$f(y)=0$ has two solutions
$0< y^\ast_1 < y^\ast_2 < 1$. In the deterministic dynamics driven by 
the bias $T_i^+-T_i^-$,
$y^\ast_1$ and $y^\ast_2$ are stable and unstable, respectively.
Therefore, case (iii) occurs.
\item If $D\le 0$, we  obtain
$f(y) \geq 0$ for all $0\le y<1$, where the equality holds true only when $D=0$ and $y=\tilde{y}$. Therefore, case (i) occurs.
\end{enumerate}
\end{enumerate}

\subsection{When $a-b-c+d = 0$}

The quadratic term in $f(y)$ disappears
when $a-b-c+d = 0$. The classification of the dynamics in this case coincides with that for
$a-b-c+d < 0$.

\section*{Appendix C: Derivation of $\tilde{q}(y)$}\appendixcount\label{sec:tilde q}

To derive the relationship $q_i = \left[\tilde{q}(i/N)\right]^N$, we 
write 
\begin{align}
q_i =& \exp\left(\sum_{k=1}^i \ln\frac{T_k^-}{T_k^+}\right)\notag\\
=& \exp\left\{
\sum_{k=1}^i \ln\frac{k\left[(k+M)c + (N-k-1)d\right]}
{(k+M)\left[(k+M-1)a + (N-k)b\right]}\right\}\notag\\
\approx& \exp \left\{N\int_0^y \ln\frac{y^{\prime}\left[(y^{\prime}+m)c+(1-y^{\prime})d\right]}
{(y^{\prime}+m)\left[(y^{\prime}+m)a+(1-y^{\prime})b\right]}{\rm d}y^{\prime}\right\},
\label{eq:q_i rewrite}
\end{align}
where $y =i/N$ and $y^{\prime}=k/N$.
Because the integral on the right-hand side of Eq.~\eqref{eq:q_i rewrite} is independent of $N$, we obtain
$q_i = \left[\tilde{q}(y)\right]^N$.
It should be noted that 
$\tilde{q}(0)=1$ is consistent with $q_0=1$.

\section*{Appendix D: Steepest descent method}\appendixcount\label{sec:steepest descent method}

As done in \cite{Antal2006BMB}, we use the steepest descent method to evaluate
$\sum^{N-1}_{k=i^\ast}(q_{N-1}/q_k)$ in Eq.~\eqref{eq:t_0 exponential} as follows:
\begin{align}
\sum^{N-1}_{k=i^\ast}\frac{q_{N-1}}{q_k}
\sim& \sum^{N-1}_{k=i^{\ast}}\left[\frac{\tilde{q}(1)}{\tilde{q}(k/N)}\right]^N\notag\\
\sim& N\int_{\frac{i^{\ast}}{N}}^1 \left[\frac{\tilde{q}(1)}{\tilde{q}(y^{\prime})}\right]^N {\rm d}y^{\prime}\notag\\
=& N \left[\frac{\tilde{q}(1)}{\tilde{q}(y^\ast)} \right]^N
\int_{\frac{i^{\ast}}{N}}^1 \exp\left[-\frac{\ln\frac{\tilde{q}(y^{\prime})}{\tilde{q}(y^\ast)}}{\frac{1}{N}}\right] {\rm d}y^{\prime}
\label{eq:sum q_k for steepest descent method}
\end{align}
where $\tilde{q}(y^\ast) = \min_{0\le y <1}\tilde{q}(y)$. We
approximate the integral by a Gaussian integral to obtain
\begin{equation}
\int \exp\left[-\frac{F(y^{\prime})}{\lambda}\right] {\rm d}y^{\prime} \approx \sqrt{\frac{2\pi\lambda}{F^{\prime \prime}(y^\ast)}}\exp\left[-\frac{F(y^\ast)}{\lambda}\right]
\end{equation}
with $F(y^{\prime})=\ln\left[\tilde{q}(y^{\prime})/\tilde{q}(y^\ast)\right]$ and $\lambda=1/N$ such that
\begin{equation}
\sum^{N-1}_{k=k^\ast}\frac{q_{N-1}}{q_k} \sim \sqrt{N} \left[\frac{\tilde{q}(1)}{\tilde{q}(y^\ast)} \right]^N.
\end{equation}

\section*{Appendix E: Weak selection introduced via an exponential function leads to fast fixation}\appendixcount\label{sec:proof with beta}

Assume that the fitness of an $A$ and $B$ player is given by
Eqs.~\eqref{eq:fi beta} and \eqref{eq:gi beta}, respectively. Then, we obtain
\begin{align}
\frac{T_i^-}{T_i^+} =& \frac{i g_i}{(i+M)f_i}\notag\\
=& \exp \left\{\beta \frac{\left[(i+M)c + (N-i-1)d\right]
- \left[(i+M-1)a + (N-i)b\right]}{N+M-1} 
+ \ln\frac{i}{i+M} \right\}.
\end{align}
If $T_i^-/T_i^+ < 1$, i.e.,
\begin{equation}
\beta \frac{\left[(i+M)c + (N-i-1)d\right]
- \left[(i+M-1)a + (N-i)b\right]}{N+M-1} 
+ \ln\frac{i}{i+M} < 0
\label{eq:cnd for fast fixation beta 1}
\end{equation}
holds true for any $i$ ($1\le i\le N-1$) and $N$,
the fixation occurs fast (i.e., $t_0\propto N\ln N$). By substituting
$y=i/N$ and $m=M/N$ in Eq.~\eqref{eq:cnd for fast fixation beta 1} and ignoring $O(N^{-1})$ terms,
we obtain
\begin{equation}
\beta \frac{(x+m)(c-a)+(1-x)(d-b)}{1+m} < \ln \frac{x+m}{x}.
\label{eq:cnd for fast fixation beta 2}
\end{equation}
Because the right-hand side of Eq.~\eqref{eq:cnd for fast fixation beta 2} is positive, there exists $\beta_{\rm c}>0$ (including the case $\beta_{\rm c}=\infty$) such that $t_0\propto N\ln N$ when $0\le \beta<\beta_{\rm c}$. It should be noted that, in contrast to the assumption throughout the present article, $a$, $b$, $c$, and $d$ are allowed to be negative in the present analysis because $f_i$ and $g_i$ given by Eqs.~\eqref{eq:fi beta} and \eqref{eq:gi beta} are positive irrespective of the $a$, $b$, $c$, and $d$ values.

\section*{Acknowledgements}

We thank Bin Wu for carefully reading the manuscript. NM
acknowledges the support provided through Grants-in-Aid for Scientific Research (No. 23681033) from MEXT, Japan, the Nakajima Foundation, CREST JST,
and the Aihara Innovative Mathematical
Modelling Project, the Japan Society for the Promotion of Science
(JSPS) through the ``Funding Program for World-Leading Innovative R\&D
on Science and Technology (FIRST Program),'' initiated by the Council
for Science and Technology Policy (CSTP).

\newpage
\clearpage

\begin{table}
\begin{center}
\caption{Classification of the three cases of the mean fixation time when $N$ is large.}
\begin{tabular}{|c|p{3.2cm}|p{3.2cm}|}
\hline
 & \hfill $a-b-c+d\le 0$ \hfill \ & \hfill $a-b-c+d>0$ \hfill \ \\ \hline
$c<(m+1)a$ & \hfill case (i) \hfill \ & \hfill case (i) or (iii)\hfill \ \\ \hline
$c>(m+1)a$ & \hfill case (ii) \hfill \ & \hfill case (ii)\hfill \ \\ \hline
\end{tabular}
\label{tab:cases 1-3}
\end{center}
\end{table}

\newpage
\clearpage

\clearpage

\clearpage

\begin{figure}[htbp]
 \begin{center}
  \includegraphics[width=15cm]{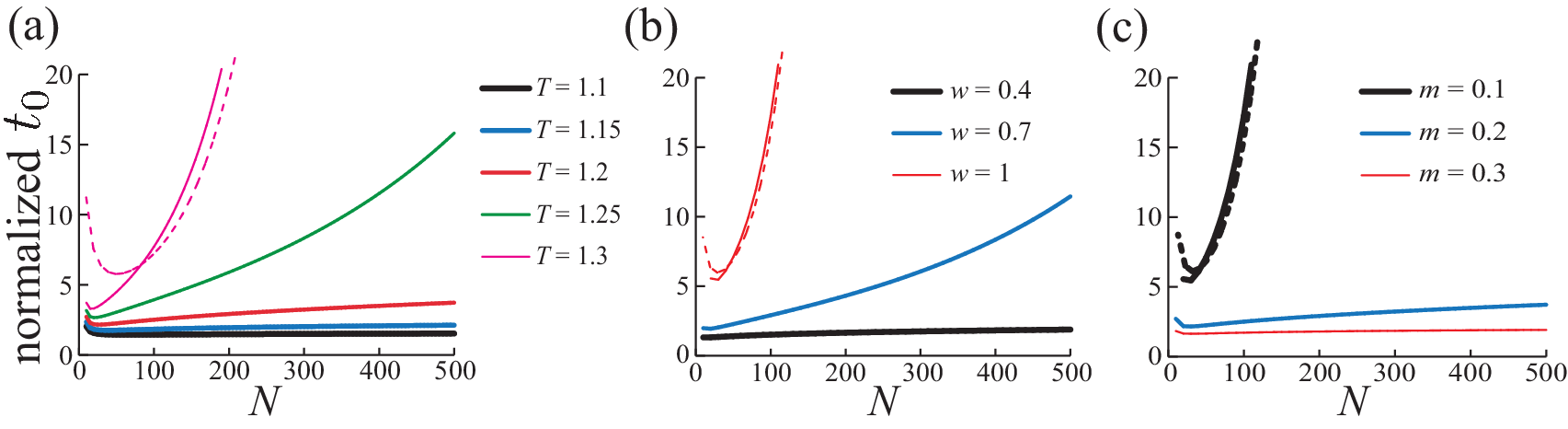}
 \end{center}
 \caption{The normalized mean fixation time for the prisoner's dilemma game as a function of $N$. We set $a=1$, $b=0$, $c=T$, and $d=0$. In (a), we set $m=0.2$ and $w=1$. In (b), we set $T=1.2$ and $m=0.1$. In (c), we set $T=1.2$ and $w=1$.
The dashed lines represent $400\sqrt{N}\exp(\gamma N)$ divided by the $t_0$ value for the neutral game.}
\label{fig:t_0 PD}
\end{figure}

\clearpage

\begin{thebibliography}{10}

\bibitem{Mobilia2003PhysRevLett}
M.~Mobilia.
\newblock {Does a single zealot affect an infinite group of voters?}
\newblock {\em Phys. Rev. Lett.}, 91:028701, 2003.

\bibitem{GalamJacobs2007PhysicaA}
S.~Galam and F.~Jacobs.
\newblock {The role of inflexible minorities in the breaking of democratic
  opinion dynamics}.
\newblock {\em Physica A}, 381:366--376, 2007.

\bibitem{Mobilia2007JStatMech}
M.~Mobilia, A.~Petersen, and S.~Redner.
\newblock {On the role of zealotry in the voter model}.
\newblock {\em J. Stat. Mech.}, page P08029, 2007.

\bibitem{XieSreenivasan2011PRE}
J.~Xie, S.~Sreenivasan, G.~Korniss, W.~Zhang, C.~Lim, and B.~K. Szymanski.
\newblock {Social consensus through the influence of committed minorities}.
\newblock {\em Phys. Rev. E}, 84:011130, 2011.

\bibitem{SinghSreenivasan2012PRE}
P.~Singh, S.~Sreenivasan, B.~K. Szymanski, and G.~Korniss.
\newblock {Accelerating consensus on coevolving networks: The effect of
  committed individuals}.
\newblock {\em Phys. Rev. E}, 85:046104, 2012.

\bibitem{Fu2011ProcRSocB}
F.~Fu, D.~I. Rosenbloom, L.~Wang, and M.~A. Nowak.
\newblock {Imitation dynamics of vaccination behaviour on social networks.}
\newblock {\em Proc. R. Soc. B}, 278:42--49, 2011.

\bibitem{LiuWuZhang2012PhysRevE}
X.~T. Liu, Z.~X. Wu, and L.~Zhang.
\newblock {Impact of committed individuals on vaccination behavior}.
\newblock {\em Phys. Rev. E}, 86:051132, 2012.

\bibitem{Masuda2012SciRep}
N.~Masuda.
\newblock {Evolution of cooperation driven by zealots}.
\newblock {\em Sci. Rep.}, 2:646, 2012.

\bibitem{Nowak2004Nature}
M.~A. Nowak, A.~Sasaki, C.~Taylor, and D.~Fudenberg.
\newblock {Emergence of cooperation and evolutionary stability in finite
  populations}.
\newblock {\em Nature}, 428:646--650, 2004.

\bibitem{Taylor2004BullMathBiol}
C.~Taylor, D.~Fudenberg, A.~Sasaki, and M.~A. Nowak.
\newblock {Evolutionary game dynamics in finite populations}.
\newblock {\em Bull. Math. Biol.}, 66:1621--1644, 2004.

\bibitem{Nowak2006book}
M.~A. Nowak.
\newblock {\em Evolutionary Dynamics}.
\newblock The Belknap Press of Harvard University Press, MA, 2006.

\bibitem{Antal2006BMB}
T.~Antal and I.~Scheuring.
\newblock {Fixation of strategies for an evolutionary game in finite
  populations}.
\newblock {\em Bull. Math. Biol.}, 68:1923--1944, 2006.

\bibitem{Traulsen2007JTheorBiol}
A.~Traulsen, J.~M. Pacheco, and M.~A. Nowak.
\newblock {Pairwise comparison and selection temperature in evolutionary game
  dynamics}.
\newblock {\em J. Theor. Biol.}, 246:522--529, 2007.

\bibitem{Altrock2009NewJPhys}
P.~M. Altrock and A.~Traulsen.
\newblock {Fixation times in evolutionary games under weak selection}.
\newblock {\em New J. Phys.}, 11:013012, 2009.

\bibitem{Altrock2010PhysRevE}
P.~M. Altrock, C.~S. Gokhale, and A.~Traulsen.
\newblock {Stochastic slowdown in evolutionary processes}.
\newblock {\em Phys. Rev. E}, 82:011925, 2010.

\bibitem{Assaf2010JStatMech}
M.~Assaf and M.~Mobilia.
\newblock {Large fluctuations and fixation in evolutionary games}.
\newblock {\em J. Stat. Mech.}, 2010:P09009, 2010.

\bibitem{Ewens2010book}
W.~J. Ewens.
\newblock {\em Mathematical Population Genetics I. Theoretical Introduction}.
\newblock Springer, New York, 2010.

\bibitem{WuAltrock2010PhysRevE}
B.~Wu, P.~M. Altrock, L.~Wang, and A.~Traulsen.
\newblock {Universality of weak selection}.
\newblock {\em Phys. Rev. E}, 82:046106, 2010.

\bibitem{Altrock2011PlosComputBiol}
P.~M. Altrock, A.~Traulsen, and F.~A Reed.
\newblock {Stability properties of underdominance in finite subdivided
  populations.}
\newblock {\em PLOS Comput. Biol.}, 7:e1002260, 2011.

\bibitem{Assaf2012PhysRevLett}
M.~Assaf and M.~Mobilia.
\newblock {Metastability and anomalous fixation in evolutionary games on
  scale-free networks.}
\newblock {\em Phys. Rev. Lett.}, 109:188701, 2012.

\bibitem{Kreindler2013GamesEconBehav}
G.~E. Kreindler and H.~P. Young.
\newblock {Fast convergence in evolutionary equilibrium selection}.
\newblock {\em Games Econ. Behav.}, 80:39--67, 2013.

\bibitem{Moran1958PCPS}
P.~A.~P. Moran.
\newblock {Random processes in genetics}.
\newblock {\em Proc. Cambridge Philos. Soc.}, 54:60--71, 1958.

\bibitem{Vankampen2007book}
N.~G. Van~Kampen.
\newblock {\em Stochastic Processes in Physics and Chemistry}.
\newblock Elsevier, Netherlands, Amsterdam, 3rd edition, 2007.

\bibitem{Redner2001book}
S.~Redner.
\newblock {\em A Guide to First-passage Processes}.
\newblock Cambridge University Press, Cambridge, 2001.

\bibitem{Krapivsky2010book}
P.~L. Krapivsky, S.~Redner, and E.~Ben-Naim.
\newblock {\em A Kinetic View of Statistical Physics}.
\newblock Cambridge University Press, Cambridge, 2010.

\bibitem{Mobilia2012PhysRevE}
M.~Mobilia.
\newblock {Stochastic dynamics of the prisoner's dilemma with cooperation
  facilitators}.
\newblock {\em Phys. Rev. E}, 86:011134, 2012.

\bibitem{Traulsen2008BullMathBiol}
A.~Traulsen, N.~Shoresh, and M.~A. Nowak.
\newblock {Analytical results for individual and group selection of any
  intensity}.
\newblock {\em Bull. Math. Biol.}, 70:1410--1424, 2008.

\bibitem{Maynardsmith1982book}
J.~Maynard~Smith.
\newblock {\em Evolution and the Theory of Games}.
\newblock Cambridge University Press, Cambridge, UK, 1982.

\bibitem{Sugden1986book}
R.~Sugden.
\newblock {\em The Economics of Rights, Co-operation and Welfare}.
\newblock Blackwell, New York, NY, 1986.

\bibitem{Hauert2004Nature}
C.~Hauert and M.~Doebeli.
\newblock {Spatial structure often inhibits the evolution of cooperation in the
  snowdrift game}.
\newblock {\em Nature}, 428:643--646, 2004.

\end{thebibliography}
\end{document}